\documentclass[epj]{webofc}
\usepackage[varg]{txfonts}   
\wocname{EPJ Web of Conferences}
\woctitle{ICNFP 2016}
\begin{document}
\selectlanguage{english}
\title{Recent results from the strong interactions program of NA61/SHINE}
%
%

\author{Szymon Pulawski\inst{1} \fnsep\thanks{\email{s.pulawski@cern.ch} On behalf of the NA61/Shine Collaboration}
}

\institute{Institute of Physics, University of Silesia, Katowice, Poland
}

\abstract{%
  The NA61/SHINE experiment studies hadron production in hadron+hadron,
hadron+nucleus and nucleus+nucleus collisions. The strong interactions program has two main
purposes: study the properties of the onset of deconfinement and search for the signatures of the critical point of strongly interacting matter.
This aim is pursued by performing a two-dimensional scan of the phase
diagram by varying the energy/momentum (13A-158A GeV/c) and the system size
(p+p, Be+Be, Ar+Sc, Xe+La) of the collisions.
This publication reviews recent results from p+p, Be+Be and Ar+Sc
interactions. Measured particle spectra are discussed and compared to
NA49 results from Pb+Pb collisions. The results illustrate the
progress towards scanning the phase diagram of strongly interacting matter.

}
\maketitle
\section{The NA61/SHINE facility}
\label{detector}
\begin{figure}[h]
\centering
\includegraphics[width=0.7\textwidth,clip]{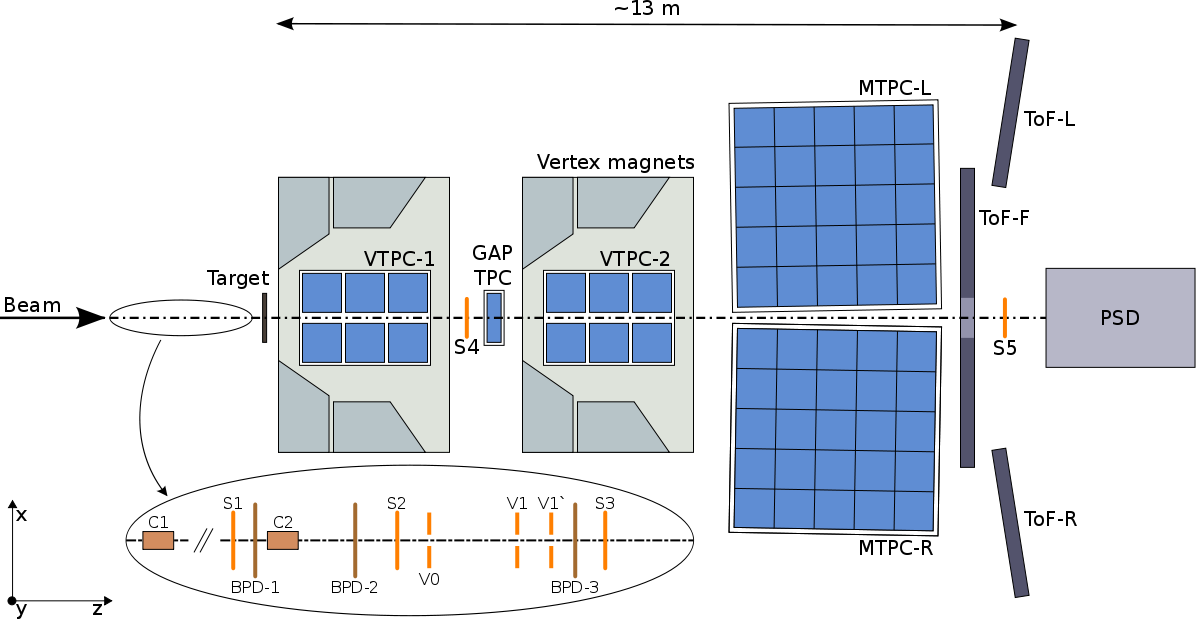}
\caption{Schematic view of the NA61/SHINE detector system.}
\label{fig:det}       
\end{figure}

The layout of the NA61/SHINE detector~\cite{detectorpaper} is presented in Fig.~\ref{fig:det}. It
consists of a large acceptance hadron spectrometer with excellent capabilities in charged particle momentum measurements and identification by a
set of five Time Projection Chambers as well as Time-of-Flight detectors.
The high resolution forward calorimeter, the Projectile Spectator Detector,
measures energy flow around the beam direction, which in nucleus-nucleus
reactions is primarily a measure of the number of spectator (non-interacted)
nucleons and is thus related to the centrality of the collision. 
An array of beam detectors identifies beam particles and measures precisely their trajectories.
Primary and secondary hadron as well as ion beams are used by the experiment.

\section{Particle ratios in inelastic p+p collisions and inverse slope parameter~T}
\begin{figure}[h]
\centerline{%
\includegraphics[width=0.6\textwidth]{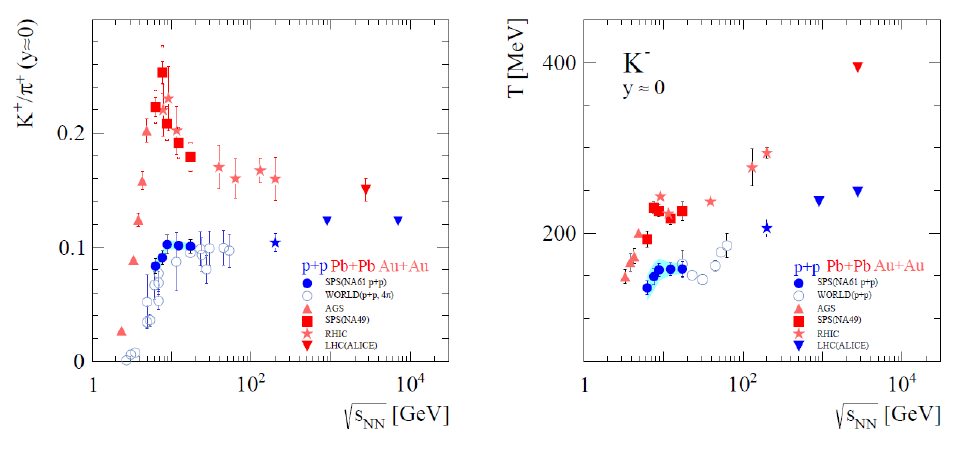}}
\caption{Energy dependence of the $K^+/\pi^+$ ratio (left) and of the inverse slope parameter T of $K^{-}$ transverse mass spectra (right) in p+p interactions compared to Pb+Pb and Au+Au collisions.}
\label{fig:horn}
\end{figure}

The excellent particle identification based on the Time-off-Flight (ToF) and energy loss (dE/dx) measurements allows to calculate
the ratio of yields $K^{+}/\pi^+$. The energy dependence of the $K^+/\pi^+$ ratio at midrapidity
for inelastic p+p interactions and central Pb+Pb/Au+Au collisions is presented in Fig.~\ref{fig:horn}~(left).
The NA61/SHINE data suggest that even in inelastic p+p interactions the energy dependence of the $K^+/\pi^+$ ratio exhibits rapid changes in the SPS energy range. However, the horn structure~\cite{6} observed in central Pb+Pb collisions is reduced to a step. Data obtained beyond the SPS energy range, namely at RHIC and LHC~\cite{7,8,9,10,11}, continue the trend seen at the SPS.

The inverse slope parameters T for $\pi$, K and p of the spectra of transverse mass $m_T=\sqrt{p_T^2+m^2}$ were fitted in inelastic p+p reactions.
Fig.~\ref{fig:horn}~(right) presents the energy dependence of T for K$^-$ in p+p collision. The NA61/SHINE
results from inelastic p+p collisions exhibit rapid changes like those observed in central Pb+Pb interactions.
World data for p+p and Pb+Pb/Au+Au reactions are plotted for a comparison and were taken
from Refs.~\cite{10,12,13,14}.

\begin{figure}[h]
\centerline{%
\includegraphics[width=0.6\textwidth]{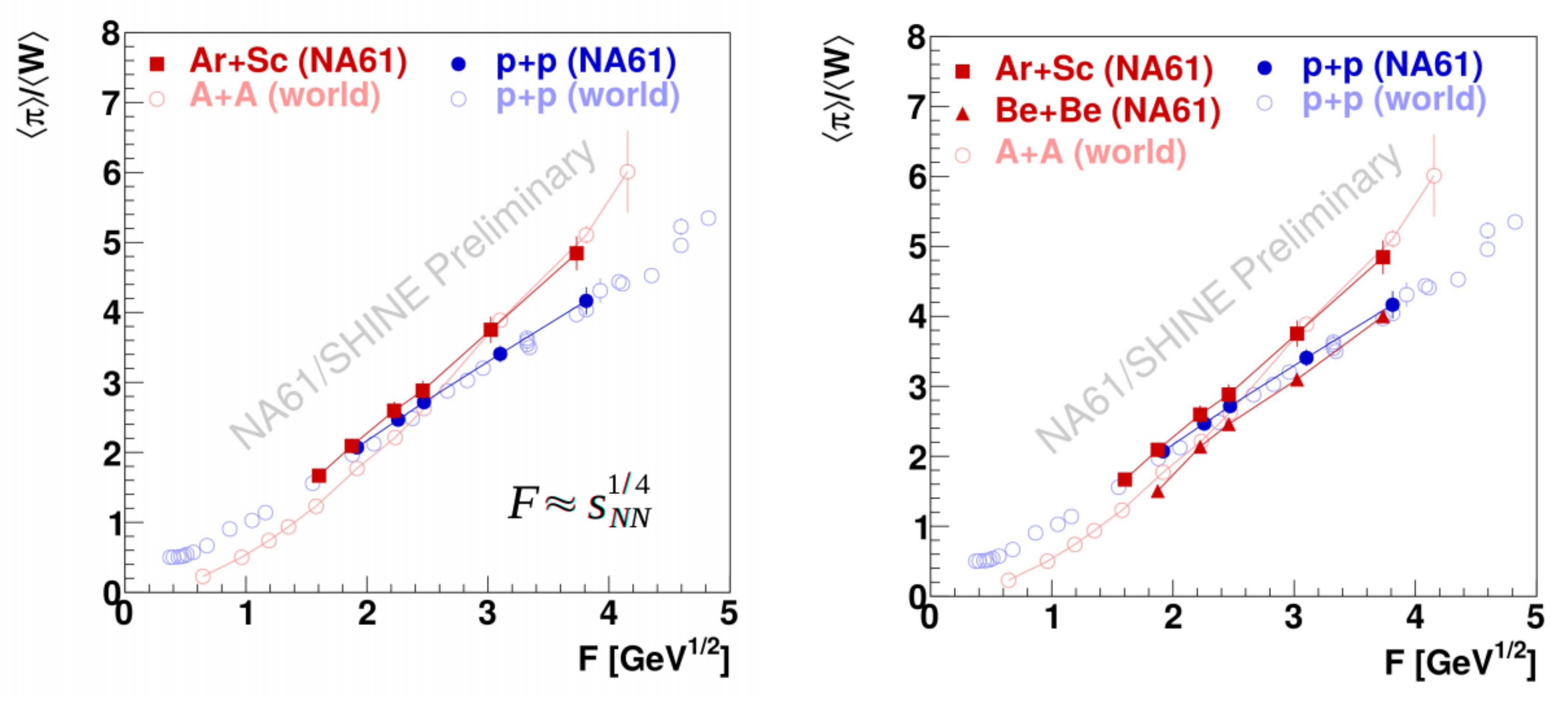}}
\caption{Mean pion multiplicity divided by the mean number of wounded nucleons as function of collision energy, 
the so called kink plot, for inelastic p+p, Ar+Sc and A+A interactions. The Fermi variable $F \approx s_{NN}^{0.25}$
was used as energy measure.}
\label{fig:kink}
\end{figure}

The dependence of the mean pion multiplicity divided by the mean number of wounded nucleons (participants) 
as function of collision energy, the so called kink plot, is shown for inelastic p+p, Ar+Sc and A+A interactions 
in Fig.~\ref{fig:kink}. The pion multiplicity $\left\langle \pi \right\rangle$ in the SPS energy range increases 
faster in central Pb+Pb than in p+p collisions. The two dependencies cross at about 40\textit{A}~GeV/c. 
For high SPS energies results from Ar+Sc reactions follow the Pb+Pb trend while for low SPS energies Ar+Sc 
follows the p+p tendency. 
The situation is opposite for Be+Be collisions.

\begin{figure}[h]
\centerline{%
\includegraphics[width=0.8\textwidth]{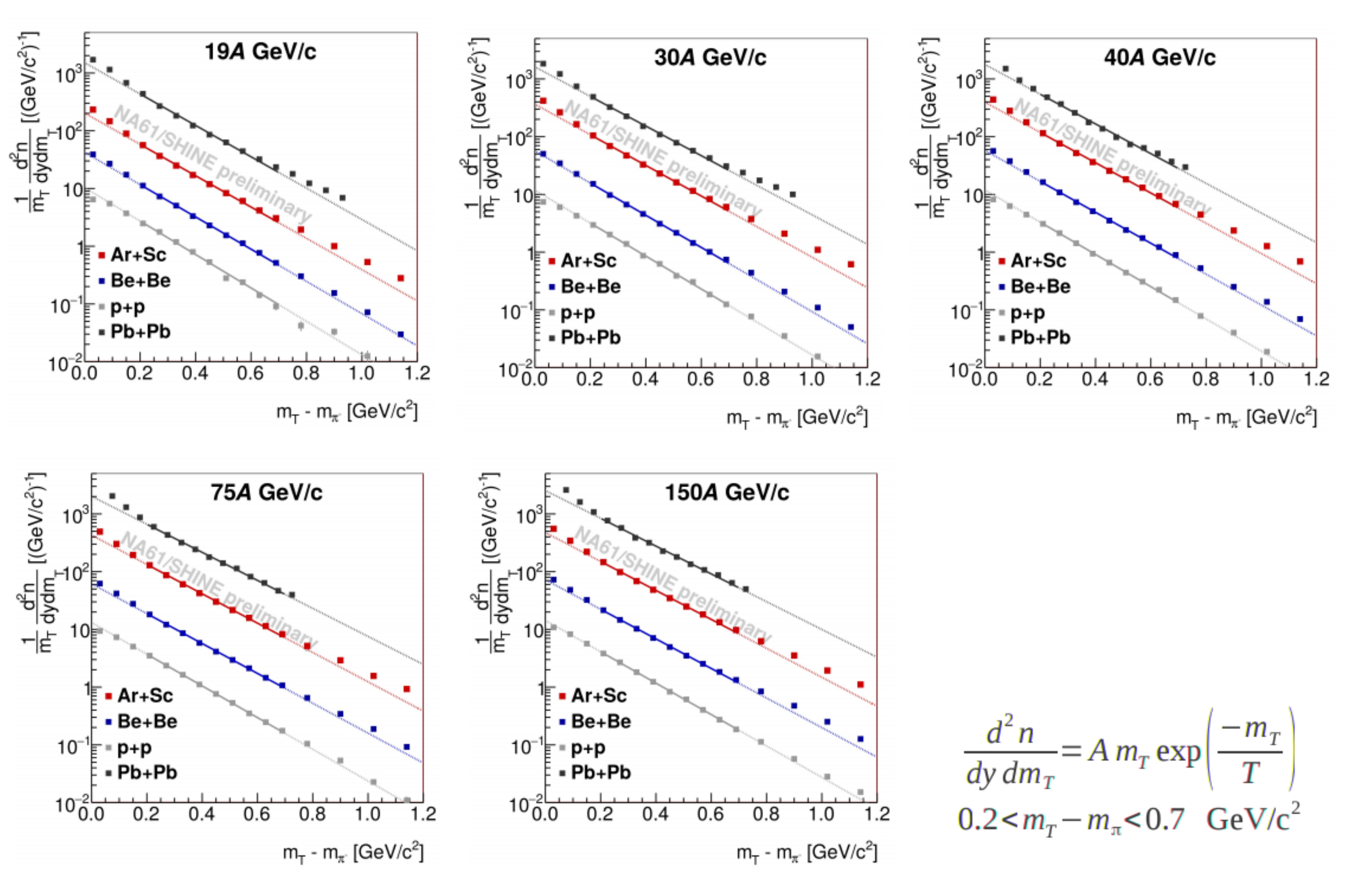}}
\caption{Spectra of transverse mass $m_{T} - m_{0}$ of $\pi^{-}$ mesons in Ar+Sc compared to 
p+p, Be+Be, and Pb+Pb collisions.}
\label{fig:mt_pim}
\end{figure}

Spectra of transverse mass $m_{T} - m_{0}$ of $\pi^{-}$ mesons in Ar+Sc are shown and compared 
to measurements for p+p, Be+Be, and Pb+Pb collisions 
in Fig.~\ref{fig:mt_pim}. Spectra in p+p reactions are exponential. A concave shape 
is observed in Pb+Pb, Ar+Sc and Be+Be collisions. The shape difference between p+p and A+A reactions
is attributed to transverse collective flow in A+A.

\section{$\Lambda$ spectra in p+p interactions}
\begin{figure}[h]
\centerline{%
\includegraphics[width=0.6\textwidth]{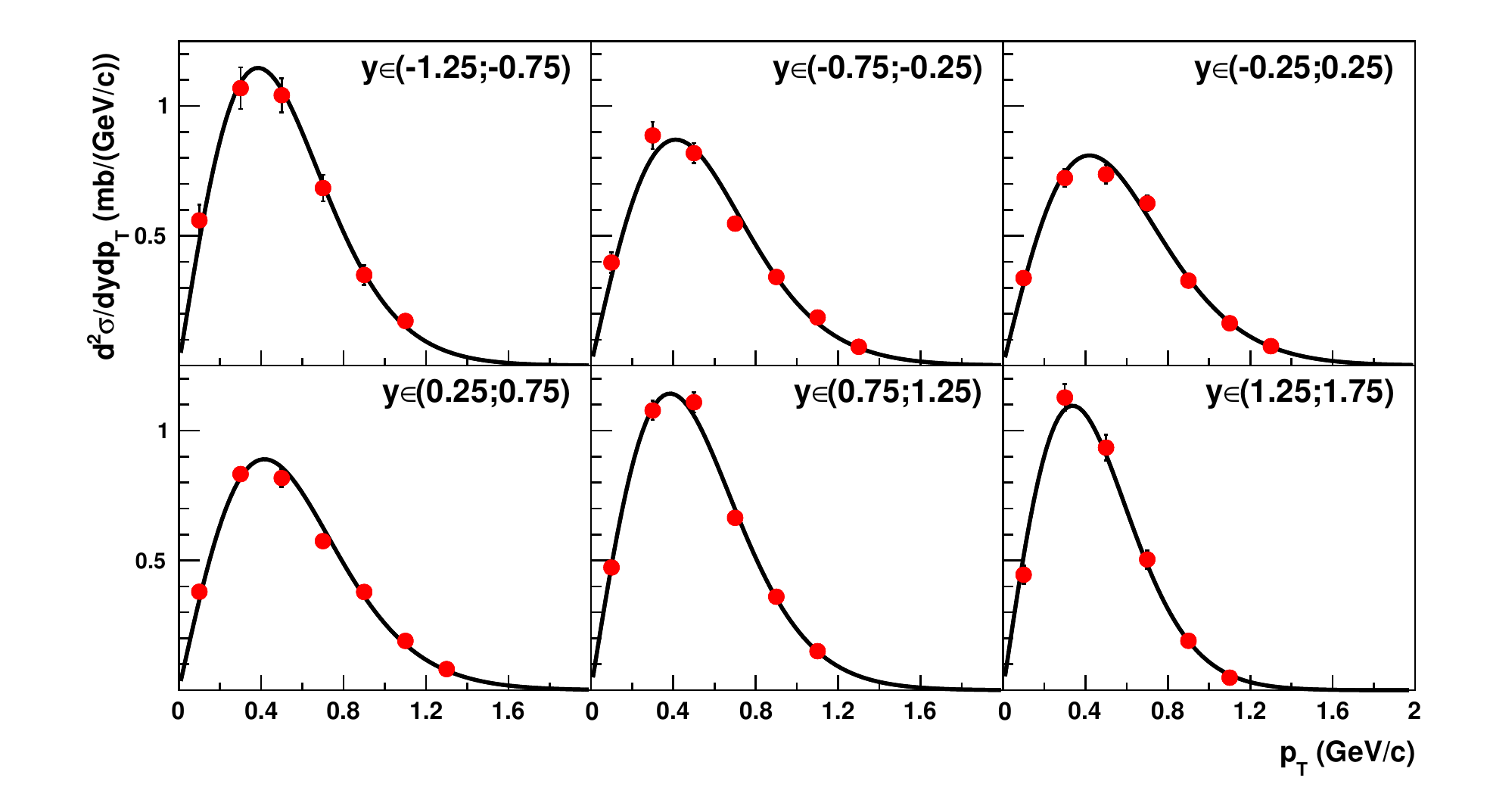} \\
\includegraphics[width=0.45\textwidth]{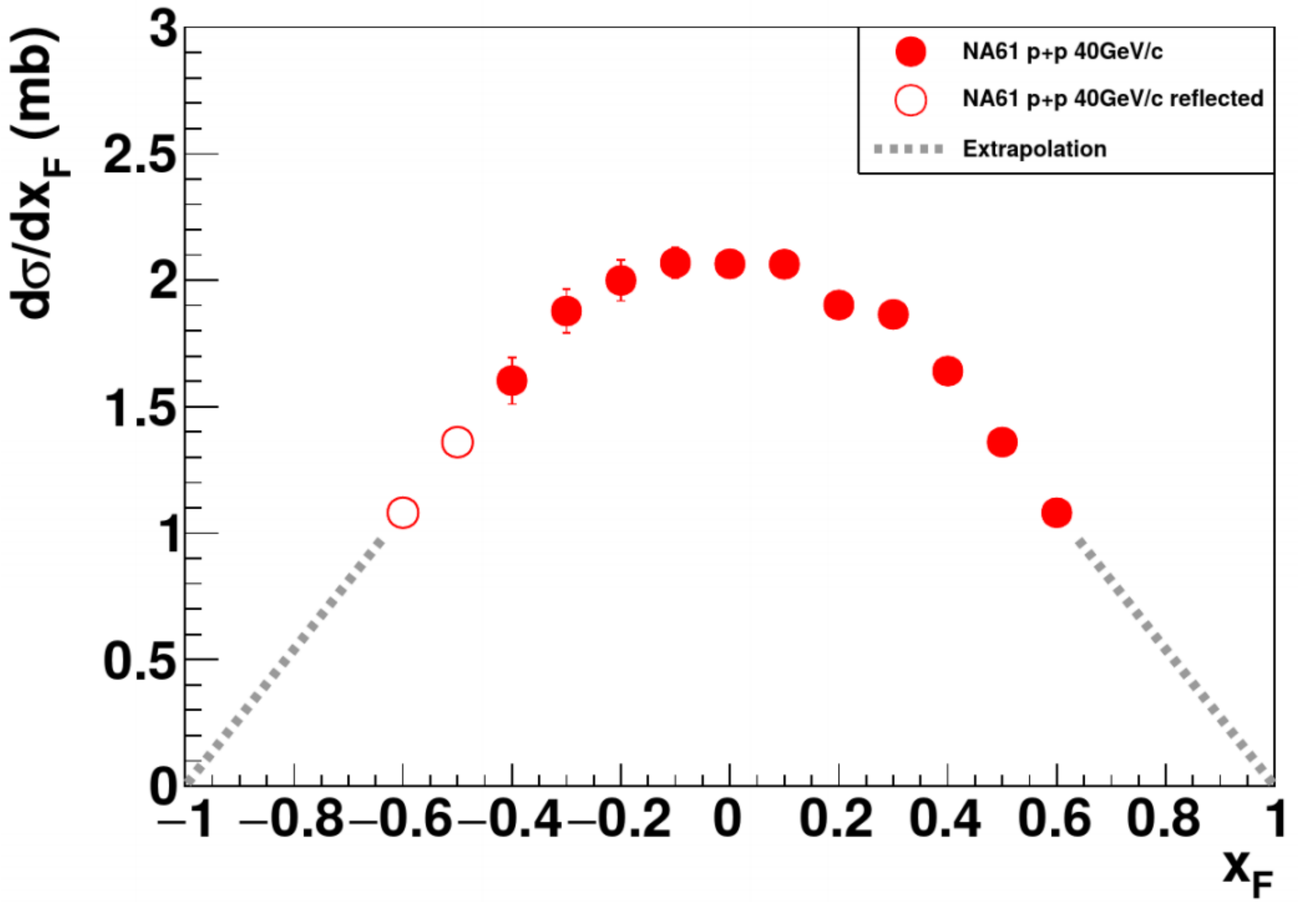}}
\caption{$\Lambda$  production cross section in p+p interactions at 40 GeV/c beam momentum.
Left: $p_T$ distributions in rapidity intervals. The lines are the result of fits to the experimental points.
Right: $p_T$ integrated cross-section as function of $x_F$. The dotted line was used to extrapolate to the
total $\Lambda$ yield.}
\label{fig:lam40}
\end{figure}

 The transverse momentum distributions of $\Lambda$ hyperons shown in Fig.~\ref{fig:lam40}~(left) are
for inelastic p+p collisions at a beam momentum of 40~GeV/c. The yields are corrected for acceptance 
and for losses due to the topological and track selection cuts.
Integration and extrapolation of the transverse momentum distributions 
provide the data points for the distributions of rapidity $dn/dy$ and the Feynman variable $dn/dx_{F}$.
The Feynman variable distribution of $\Lambda$ hyperons produced in inelastic p+p reactions at 40~GeV/c is
shown in Fig.\ref{fig:lam40}~(right).  
%

The $\Lambda$ cross-section produced in inelastic p+p interactions at 40~GeV is presented 
in Fig.~\ref{fig:lam40}~(left). Results are shown as function of transverse momentum in rapidity intervals. 
Fig.~\ref{fig:lam40}~(right) depicts the dependence of the $\Lambda$ production cross-section as function 
of $x_{F}$ together with the extrapolation line for determining the total yield.

\begin{figure}[h]
\centerline{%
\includegraphics[width=0.5\textwidth]{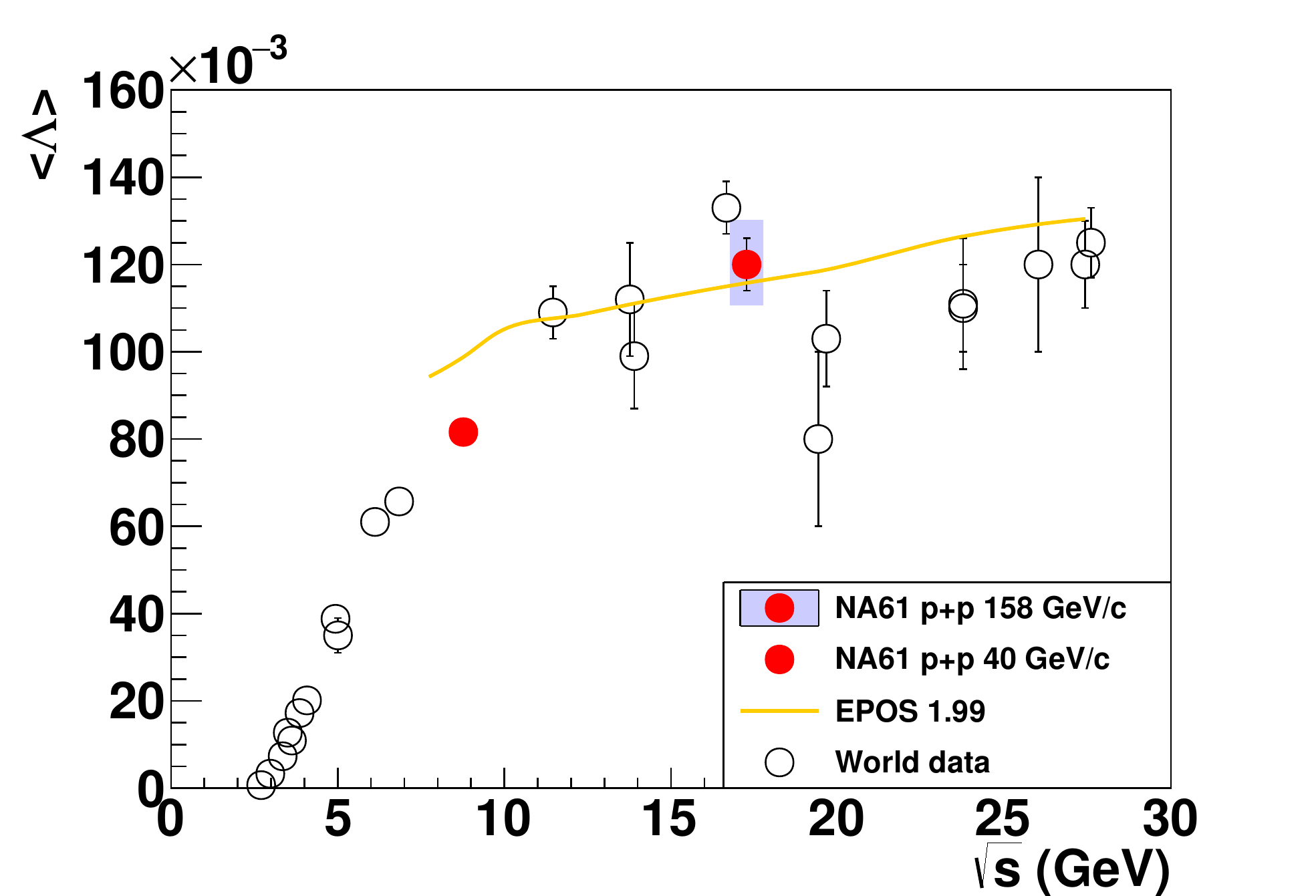}
\includegraphics[width=0.4\textwidth]{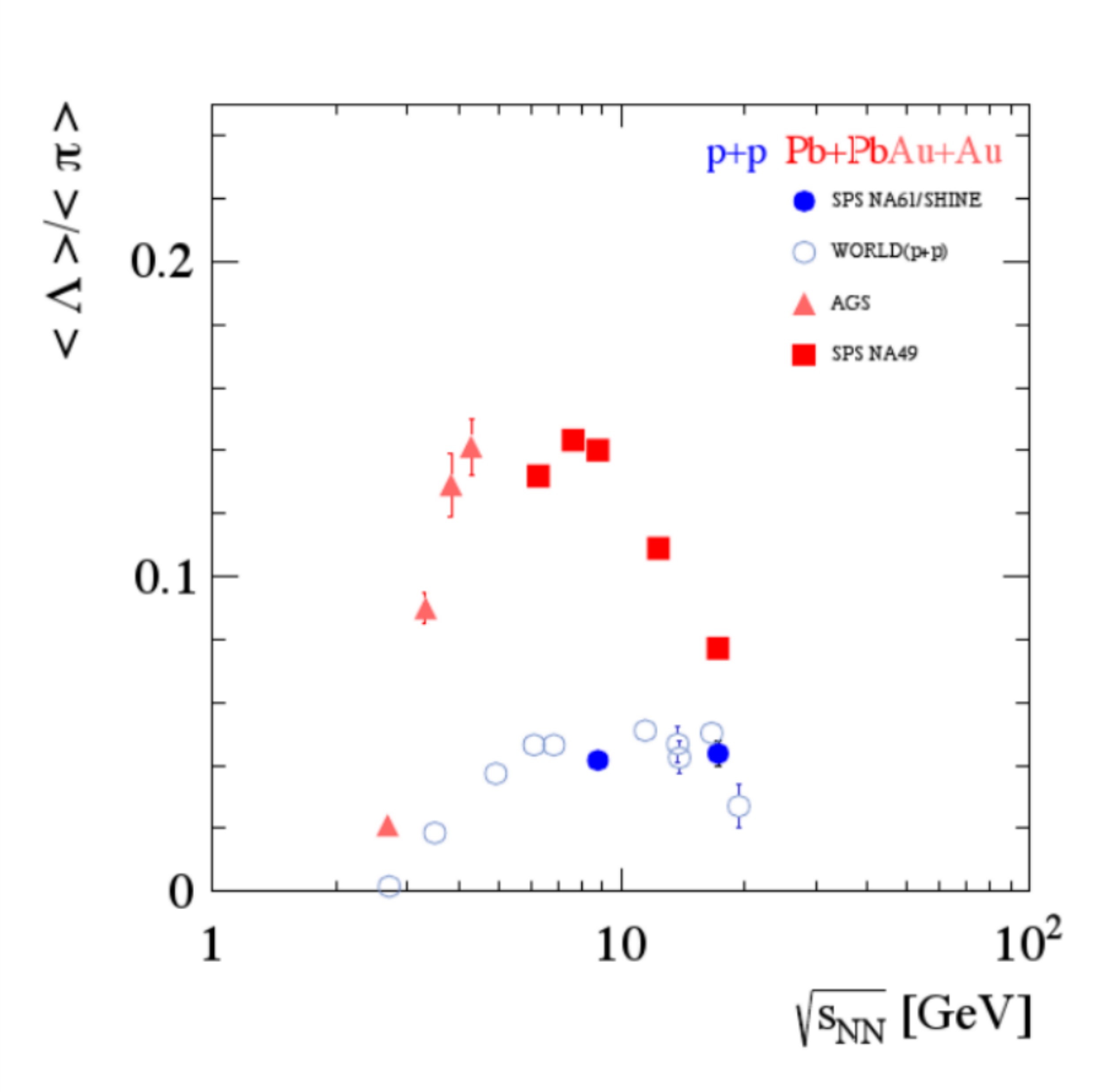}}
\caption{Left: $\Lambda$ multiplicity in p+p interactions as function of $\sqrt{s_{NN}}$ together 
with world data and results from calculations with the EPOS model. Right: mean multiplicity of 
$\Lambda$ hyperons divided by the mean $\pi$ multiplicity.}
\label{fig:lamS}
\end{figure}

Extrapolation of the $x_F$ distributions into the unmeasured region
provides estimates of the $\Lambda$ yields. The corresponding multiplicities (average yields per event)
are compared to world data in Fig.~\ref{fig:lamS}~(left) as function of center-of-mass energy $\sqrt{s_{NN}}$. 
The NA61 data follow the trend set by the world data and at the same time reduce the uncertainties of the
yields significantly at the two energies analysed by NA61. Above $\sqrt{s_{NN}}=19$ GeV the experimental 
data have large
errors and lie consistently below the EPOS model calculation. It is desirable to obtain new experimental
data in this energy range. Below 10 GeV EPOS seems to overpredict $\Lambda$ yields. However,
this model is not meant to provide reliable results at this energy and below. The ratio $\left\langle \Lambda \right\rangle / \left\langle\pi\right\rangle$ 
is shown in Fig.~\ref{fig:lamS}~(right) as function of $\sqrt{s_{NN}}$. The 
presented NA61/SHINE results agree with the world data. Measurements in A+A collisions at the AGS 
and by NA49 show a different behaviour than that observed in p+p reactions.

\section*{Acknowledgments}
This work was supported by the Polish Ministry of Science
and Higher Education (grants 667\slash N-CERN\slash2010\slash0,
NN\,202\,48\,4339 and NN\,202\,23\,1837), the Polish National Center
for Science (grants~2013\slash11\slash N\slash ST2\slash03879,
2014\slash12\slash T\slash ST2\slash00692, 
and
2015\slash18\slash M\slash ST2\slash00125),


%
%
%


\end{document}